# The Effect of Multiple Internal Representations on Context Rich Instruction

N. Lasry & M.W. Aulls

Abstract: This paper presents *n-coding,* a theoretical model of multiple internal mental representations. The *n-coding* construct is developed from a review of cognitive and imaging studies suggesting the independence of information processing along different modalities: verbal, visual, kinesthetic, social, etc. A study testing the effectiveness of the *n-coding* construct in classrooms is presented. Four sections differing in the level of *n-coding* opportunities were compared. Besides a traditional instruction section used as a control group, each of the remaining three treatment sections were given context rich problems following the "cooperative group problem solving" approach which differed by the level of *n-coding* opportunities designed into their laboratory environment. To measure the effectiveness of the construct, problem solving skills were assessed as was conceptual learning using the Force Concept Inventory. However, a number of new measures taking into account students' confidence in concepts were developed to complete the picture of student learning. Results suggest that using the developed *n-coding* construct to design context rich environments can generate learning gains in problem solving, conceptual knowledge and concept-confidence.

In his Nobel lecture, Richard Feynman stated: "*Perhaps a thing is simple if you can describe it fully in several different ways without immediately knowing that you are describing the same thing*"[1]. The converse may also be true as the ability to represent an object or phenomenon in multiple ways may simplify one's understanding of it. Consistent with this idea, several studies have addressed the importance of multiple representations in learning physics[2,3]. However, previous studies have defined representations primarily as ***external*** representations used in practice such as mathematical, diagrammatic or graphical representations[2]. The present paper shifts the focus to ***internal*** mental representations to provide insights into optimizing student learning and potentially explain the effectiveness of external representations. Just as multiple *external* representations can enhance problem solving[4], previous cognitive studies have shown that the construction of multiple *mental* representations can also enhance problem solving abilities[5]. Multiple mental representations complete each other, resulting in a more authentic portrayal of a problem than any single source of uni-modal information[6].

The present paper has three parts. The first consists of a survey of the cognitive literature to build a coherent model of internal representations that is simple enough to be used in classrooms. The second part presents a physics instructional study testing the effectiveness of the proposed model of internal representations. To achieve a comprehensive picture of effectiveness, an objective of this paper is to develop new measures of learning in physics. By shifting the focus away from a Boolean view of students' conceptual states (i.e. away from a perception that "they either get it or they don't"), these new measures acknowledge the complexity of students' conceptions by taking into account affective factors such as students' confidence in concepts. The third part will discuss the findings, their limitations and provide recommendations on maximizing physics learning.

*From encoding to n-coding*

Within the brain, the mind is currently seen as a set of specialized information processors that are spatially independent but functionally interrelated[7]. Local brain areas have different processing functions and although spatially independent, these processors interact[1] with each other, a collaboration of separate entities that Minsky has poetically called the *Society of Mind*[8]. Cognitive scientist and Nobel laureate Herbert Simon had argued that this type of modular design of the mind is but a special case of modular, hierarchical design of *all* complex systems[9].

The process through which information is taken from the external environment and "coded" for the mind is called *encoding*. Encoding can take place in several modes. Consider for instance Dual-Coding Theory (DCT). The DCT approach[10] attempts to give equal weight to verbal and visual processing. The reasoning behind this approach is that both the visual and the auditory system can be activated independently although the two systems are interconnected. In the neuro-cognitive literature these independent systems are referred to as the *auditory* or *phonological loop* and the *visuospatial sketchpad*[11]. This independence can be easily demonstrated by asking subjects to perform two simultaneous tasks. If both tasks are auditory (or both visual), an interference occurs prohibiting their simultaneous completion. However, when one is visual and the other auditory, simultaneous tasks can be performed. It has been suggested that the connectedness of both systems allows individuals to cue from one system to the other which facilitates interpreting the environment[12]. The inference is that since there are two distinct ways of encoding and representing information in our mind, the use of both representations allows parallel processing to occur thus reducing computational time.

---

[1] On possible candidate for this interaction mechanism is coherent gamma wave synchrony (Crick & Koch, 1995). This process shows how different parts of the brain can be united by allowing *local* high frequency brain waves (i.e. gamma waves) to be synchronized with other parts of the brain. Corollaries of this hypothesis have also been observed as disorganized though in schizophrenia has been correlated to gamma *a*synchrony (Haig *et al*, 2000).

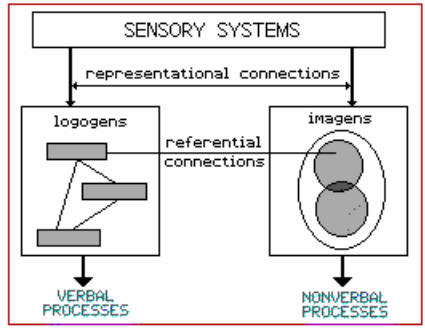

Fig 1. Reproduced with consent from TIP database

Evidence of specialized processing is abundant. Through studies of injured patients, it has been known for over a century that the processing of language is located in what are now called Broca and Wernicke's areas in the brain (see Figure 2). However, in the past 15 years, the use of neural imagery such as PET and fMRI scans has revealed an increasingly clear picture of localized processing. The impact of current advances in imaging has been likened to Santiago Ramon y Cajal's first observation of an individual nerve cell[13]. Imaging data of localized processing shows that visual and auditory words activate (a largely left sided) set of areas of the anterior and posterior cortex and the cerebellum; while simple arithmetical tasks activate left and right occipital and parietal areas[14]. There is now also information on spatial tasks[15], on understanding of the minds of others[16] and of oneself[17] and even on the processing of musical tasks[18]. Thus, encoding information from the environment is a parallel process: one part of the brain does not wait in sequence to start processing if another part is activated; more than one part can be activated at once. This survey of neuro-imaging studies suggests that multiple forms of encoding can take place. In a similar way that dual-coding theory urged us to consider two encoding modes, these imaging findings urge us to *reconsider encoding as n-coding.* The term *n-coding* is coined here to emphasize that 'n' is no longer two (as in *dual* coding) but the number of encoding dimensions identifiable, which may increase with further imaging studies.

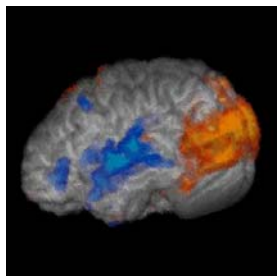

Fig 2. Visual processing areas in orangeverbal processing areas in blue

*N-coding* modes need not be strictly perceptual. For instance, imaging data on understanding the minds of others[16] supports cognitive "*theories of mind*" positing the existence of an internal "mind reading system"[19]. This modality can be seen as rooted both in genetic and environmental settings as illustrated by its dysfunction in autistic disorders[19] and its normal developmental trajectory in children[20]. However, encoding information about the mind of others is done somewhat independently from its related visual perceptual process. Indeed, autistic individuals can recognize people without being able to tell their emotional state; something autistic professor of animal studies Temple Grandin has likened to being as an anthropologist on Mars[21], constantly trying to figure out how different behaviors translate into emotional states. In a nutshell, the possibility to *n-code* is the possibility to represent information mentally along multiple dimensions (verbal, logico-mathematical, visual, kinesthetic, social, etc.). The remaining question is: How can these findings be applied to instruction?

*From theory to Practice*

It is always necessary to err on the side of caution when going from a "*descriptive learning theory*" to a "*prescriptive instructional theory*"[22]. It would be ill-advised to take imaging findings and make clear conclusions and recommendations as to what instructional environments should be like. Current educational trends, such as "Brain-based" learning, have been severely critiqued[23] for inferring one-to-one relationships between fundamental research findings and instructional practices. Before jumping into tentative applications, one is reminded of the wise comments of a founding father of American psychology, William James[24]:

> "*You make a great, a very great mistake if you think that psychology, being a science of the mind's laws, is something from which you can deduce definite programmes and schemes and methods of instruction for immediate schoolroom use*"

Although one must be wary of translating neuroscience findings directly into educational practice, it would be unconscionable not to use neuroscience findings as a guide for empirically driven educational development. The current paper will thus present a classroom quasi-experiment where students in different sections were assigned to activities based on context rich problems consistent with the Cooperative Group Problem Solving (CGPS) approach[25] but that varied in *n-coding* opportunities. The CGPS approach was chosen because of its demonstrated effectiveness[26] and because it constitutes an exemplary Interactive Engagement approach using context rich problems. The classroom experiment presented attempts to test whether increasing *n-coding* opportunities in CGPS has even more beneficial effects on learning.

*Description of treatment conditions*

Students enrolled in an algebra-based mechanics course in a Canadian 2-yr community college were given homomorphic CGPS problems. The classroom study consisted of 4 sections differing by the level of *n-coding* built into the environment and the problem structure. To get a better picture of the difference between these treatment sections it may be useful to briefly describe one CGPS activity and its implementation across the different groups. One problem was based on a popular TV show (CSI: *Crime Scene Investigators*) and put the student in the shoes of a detective trying to solve a murder. The objective of this ballistics problem was to help students better acquire and synthesize notions of 2D kinematics. The first section, labelled high *n-coding* group (nCodHi), incorporated multiple representations in the problem **presentation** and required *n-coding* in the problem **solution** as well. Thus, in the first treatment section (nCodHi), the text problems were accompanied by rich visuo-spatial presentations. Besides diagrams, an actual scene reconstitution included an outline of a victim taped on floor and chalk dust emulating gun powder distribution so as to locate the initial horizontal position of the gun. Furthermore, kinesthetic data was also available such as a bloc of wood in which a stray bullet was found and an actual 9mm slug (graciously provided by the police technology program of the college). Students were then given a table of muzzle speeds (i.e the initial speed of the bullet as it exits the gun) for various calibres. Once extracted from the bloc of wood it had supposedly been shot in, the slug's angle of entry was to be measured (a 9+mm hole was drilled at $5^\circ$ from horizontal and the slug was inserted prior to lab). Using all the information available students had to collaboratively determine which variable they were ultimately looking for. In this particular case, the initial vertical position of the bullet was to be found so as to find the approximate shoulder height of the shooter to later identify him from a line-up.

The second section, labelled medium *n-coding* group (nCodMed), did not include emulated environments that required elaborate setups. Problems were presented in text format. However, the **solution** of these problems required students to inquire about the objects of the emulated environment and make some measurements. Thus, the second group (nCodMed) was presented neither with an emulated environment nor with physical props. However, to solve the problem students had to enquire about the props. For instance, the actual 9mm slug was not presented but was available on student request and its measurement was essential to the problem solution (without the information that this was a 9mm slug, student could not know for instance the initial speed of the bullet). Thus, although the same problem was given and similar measurements had to be carried through in both groups, only one group had the props presented with the problem while the other had to enquire about them. Note that the lab problems for the two first treatment conditions were designed to use measurement instruments that were familiar to students (everyday tools such as a bathroom scale or a measuring tape, stop watches etc.).

The third section, labelled low *n-coding* group (nCodLo), was comprised of conventional CGPS context rich problems presented in text format without transforming the environment. This section had less opportunities for multiple representations as none of the props were available and all data gleaned along visual or tactile dimensions were obtainable in text form. For instance, the problem stated: "a 9mm slug was recovered from the scene". Given the calibre of the bullet the students can figure out its muzzle speed from a table. As in the other two treatment conditions, all the information was not available in the initial problem description. However, students could not measure the missing information from various props.

Finally, the control group was comprised of traditional highly-structured "cookbook" labs[27] assessing the same learning outcomes (eg. 2D kinematics). All sections of the course (treatments and control) had the same instructor (N.L.) so as to minimize, if not control for inter-instructor differences and macro-differences in classroom culture. Furthermore, all three treatment sections required students to collaboratively solve problems in groups of three or four. Students in the control section however were assigned to groups of two. The table below summarizes the presence or absence of characteristics across all groups.

**Table 1: Schema of various treatment sections**

|  | Control | nCodLo | nCodMed | nCodHi |
|---|---|---|---|---|
| Cooperative group | N | **Y** | **Y** | **Y** |
| *n-coding*/multiple representation required for ***task solution*** | N | N | **Y** | **Y** |
| *n-coding* in ***task presentation*** and ***task solution*** | N | N | N | **Y** |

*Measures*

Physics understanding is traditionally measured through procedural problem solving. In this study, these skills were assessed using the local physics department's comprehensive final examination. This exam was constructed by a committee of physics professors and had to be approved unanimously by all those teaching the course (10-12 instructors). The exam score was also consistent as each instructor marked a single exam question for the entire cohort (not just for his or her students). This insured that no group had an exam of a differing difficulty, nor a corrector of different generosity. Furthermore, the correctors of the exam questions were unaware of which students belonged to which treatment condition.

In physics, students may know how to solve problems without having a complete conceptual understanding of the physics involved[28]. Therefore, conceptual understanding was also measured. Students were assessed the first and last week of the semester with the Force Concept Inventory[29] "*the most widely used and thoroughly tested assessment instrument*" in physics[30]. To get a reliable indicator of conceptual learning, Hake's normalized FCI gain (***g***) was used[26]. The FCI normalized gain is defined as a ratio of the increase in correct concepts acquired after instruction to the maximal possible increase:

**g = (Post T – Pre T)/ (max T – Pre T)          Eq.1**

The term T in the previous equation refers to the Test score, in this case the FCI score. This normalized gain may be computed by student and the average student gain <g> can be found. Alternately, it is possible to compute this gain from the class Pre Test average score and class Post test average yielding the average class normalized gain *g*. Differences between these 2 forms of normalized gain usually reside in the fact that the initial and final population are not identical. However, differences may exist even when both populations are exactly matched[41]. To escape the potential bias due to different populations, both gains were calculated for matched students (same student pre and post) only.

Students' conceptual models are loosely organized[31] as can be seen when instances of a concept are correctly expressed in one context and not in a somewhat different context[32]. Since students conceptions seem not to fit in Boolean true-false categories[33], a ***new*** measurement is proposed combining levels of confidence to FCI questions. In his Peer Instruction approach, Mazur has shown how students' confidence levels for in-class concept-tests vary at different test times[34]. Similarly, it may be interesting to assess students' confidence for each FCI item at the beginning and the end of a semester. The level of confidence expressed for a concept allows one to infer how strongly a conception is held. Associating a level of confidence on a 5 point Likert scale (0= guessing; 1= not sure; 2= pretty sure; 3=confident; 4= Very confident) with each answer gives a better representation of students' conceptual state than the currently prevalent true-false view. The simple procedure of assessing confidences for FCI items yields 3 measures.

1) ***Average level of confidence***: represents the individual's overall confidence in answering conceptual physics questions. This level of confidence can be compared between both test times to determine the effect of treatment conditions on students' overall confidence regarding physics concepts. This could reveal interesting information particularly if an increase in confidence was to be found in some sections more than others. On the other hand, students may be less confident, which may occur if the new knowledge acquired is under construction and not fully "compiled"[35]. Note that pretest and postest average confidence levels can be also be used to compute a normalized average confidence gain. To find the normalized average confidence gain, the term T in equation 1 must simply be replaced by the average FCI confidence (AVGconf).

2) *Confidence level for Right/Wrong answers*: can be contrasted at both test times. For instance, are students significantly more confident of correct answers at the end of the semester? Also, are students more confident in right than wrong answers before/after instruction? Here again, confidence gains can be normalized. To find the normalized average confidence gain, the term T in equation 1 must simply be replaced by the average right FCI confidence (Rconf) or average wrong FCI confidence (Wconf).

3) *Weighted FCI* score. Assuming that a 5 point Likert scale can be treated as a continuum (which is implicitly done when researchers perform t-tests on Likert scale data for instance), we can associate a numerical value to each level of confidence and use this as a factor in determining a "weighted" FCI score. Let us attribute 1 point for a correct answer and –1 point for an incorrect answer. Levels of confidence are values corresponding to the student entry: 0 on the scale indicating "guessing" and 4 indicating "very confident". A student entering a good answer with maximum confidence gets 4 points (1 x 4) whereas a student entering a wrong answer with maximum confidence receives –4 points ( -1 x 4). Students that are not at all sure of an answer (i.e. confidence level 0) such as students that are guessing, get 0 points regardless of whether the answer is right or wrong. The 2-point true-false representation of students conceptions can now be mapped on a 9-point pseudo-continuum: from highly confident in a misconception (-4) to highly confident in a correct conception (+4). Resulting total weighted scores for the 30 FCI items therefore vary between –120 and 120. Differences in *weighted* FCI score across all groups can then be compared between both testing occasions. Here again, weighted FCI gains can be normalized by replacing T in equation 1 by the weighted FCI score (wFCI) yielding:

**$wg$ = (Pre *wFCI* – Post *wFCI* )/ (120 – Pre *wFCI*)         Eq.2**

These measures may address some of the concerns raised by the interpretation of FCI scores[36]. For instance, a student guessing a right answer would not attribute high confidence to an item. Therefore, a portion of false positives (students guessing a right answer) would become identifiable. Furthermore, these measure are more comprehensive as they assess *cognitive* changes (conceptual change) *and affective* changes (confidence change) in physics learning thus addressing 2 of the 3 fundamental components of the mental trilogy: cognition, conation (i.e. motivation) and affect[37].

Possibly the **most important** use of wFCI is as a diagnostic tool. Indeed, compiling confidence data across a question (instead of across a student) before instruction, it is possible to identify strong group misconceptions (high confidences for wrong answer). For instance, if pretest results show that the class average wFCI for a given item is highly positive (closer to +4), the instructor can then decide to briefly overview the concept and reallocate the time allotted to teaching it. On the other hand if the pretest results show that the class average wFCI for a given item is highly negative (closer to -4) instructors would be alerted to a misconception strongly held by many students and could devote more time to addressing these misconceptions. Instructors could thus adapt their course by briefly overviewing concepts that were largely and confidently understood to devote more time and effort to misconceptions that exhibit high confidence levels across the group. Such analyses will be presented elsewhere as they do not fit within the objectives of the current study. Using the proposed measures, the central research question is: does instruction using *n-coding* opportunities provide better learning gains?

*Sample*

Participants consisted of a cohort of 84 students following first semester algebra-based physics, pseudo-randomly assigned by the registrar to 4 different laboratory sections following distinct instructional formats (nCodHi: n= 20; nCodMed: n=20; N-CodLo: n=24; Cont: n=20). Instruction for all four sections of this study took place in the same laboratory, a conventional physics laboratory with seating for 24 students. Of the initial 84 students pre-tested, 61 were also post-tested (nCodMed: n=14; nCodHi: n= 14; nCodLo: n=18 ; Cont: n=15). This attrition may be explained in part to the loss of first semester students due to program changes as well as decreases in attendance in the week prior to final examinations. Data in the form of pre and post FCI scores, related confidence levels and final exam scores were collected for these 61 participants.

*Analysis*

Using a multivariate analysis of variance (MANOVA), treatment condition was taken as an independent variable and the FCI, wFCI, right answer confidence (RConf) and the final exam grade (Exam) were taken as the multiple dependent variables. These four dependent variables were chosen as each gives some information the others do not. For instance, problem solving aptitude is a measure of learning. However, students may learn how to solve problems algorithmically without full conceptual understanding[28]. The FCI therefore completes the picture by measuring conceptual learning. Yet, since answers on the FCI are either right or wrong and students' conceptions seem not to fit in Boolean categories[33] a confidence weighted FCI was developed to steer away from black and white categorizations and create a richer grayscale. Finally, since a robust use of a conceptual model entails a good confidence in correct concepts, the right answer confidence gain was also sought to complete the picture. A MANOVA is an appropriate statistical measure here as it seeks differences between groups by adding the contributions from each dependent variable (dependent variables must be linearly independent of one another). Thus, differences under one specific variable such as the FCI may not be significant between groups –particularly with small sample sizes such as those in this study - but the combined effect of the different variables may result in significant differences.

Three planned group comparisons seek differences between: 1) *n-coding* CGPS (nCodHi&Med) and conventional CGPS (nCodLo); 2) high vs medium *n-coding* groups; 3) all CGPS groups vs control group. The rationale behind these planned comparisons follows from our theoretical model. We hypothesized that increasing *n-coding* should entail increased learning gains. Since the high and medium *n-coding* groups are CGPS groups that have been modified by additional *n-coding* opportunities, the first comparison seeks differences between these *n-coding* transformed CGPS groups and conventional CGPS groups. The additional advantage of combining both the high and medium *n-coding* groups is to increase (in fact double) the small number of students in the sample and provide more power to find differences between the groups should they exist. Second, we compare the high vs medium *n-coding* groups to see whether adding *n-coding* in the *presentation* increases learning gains significantly. Finally, we group all three CGPS active engagement groups to see whether the differences expected in conceptual learning also transpire in confidence gains and confidence weighted FCI gains.

*Results*

The average value per group for each of the four dependent variables measured (exam, FCI, wFCI, Rconf) is presented in table 2. These data show high *n-coding* group outperforms every other group; the exception being a marginal difference between the medium *n-coding* and high *n-coding* groups on right answer confidence gains. These data also show that medium *n-coding* outperforms the conventional CGPS and Control sections for each variable. Finally, the conventional CGPS group did better than the Control section on all measures aside from the final exam grade. Thus it would seem that the level of *n-coding* is a good variable to hierarchically classify groups according to their learning gains. That is, the greater the *n-coding* the greater the learning gains. The remaining question is: are these differences statistically significant or are they due to chance?

**Table 2: Dependent variable results for each group**

|  | nCodHi (n=14) | nCodMed (n=14) | nCodLo (n=18) | Control (n=15) |
|---|---|---|---|---|
| *FCI g* | 0.46 ± 0.21 | 0.42 ± 0.24 | 0.32 ± 0.23 | 0.18 ± 0.29 |
| *wFCI g* | 0.33 ± 0.31 | 0.32 ± 0.28 | 0.20 ± 0.19 | 0.14 ± 0.20 |
| *RConf g* | 0.21 ± 0.36 | 0.22 ± 0.58 | 0.09 ± 0.48 | 0.03 ± 0.47 |
| *Exam* (Avg ± SD) | 83.6 ± 9.4 | 77.1 ± 11.6 | 63.3 ± 17.2 | 69.9 ± 10.8 |

In comparing the difference in these dependent variables between groups using a MANOVA, the hypothesis of "No Overall Treatment Effect" between groups was significantly rejected (p= 0.009). Since the MANOVA shows that the groups differ significantly in outcome, the question remains which groups differ? The differences in the MANOVA for the three planned comparisons between groups are presented in table 3.

**Table 3: MANOVA Planned Comparison**

|  | *p* |
|---|---|
| *nCodMed vs. nCodHi* | 0.740 |
| *nCodHi&Med vs. nCodLo* | *0.003* |
| *nCodHi&Med&Lo vs. Cont* | *0.0497* |

These data shows that the two *n-coding* CGPS groups (nCodMed and nCodHi) differed statistically from the conventional CGPS approach (p=0.003). However, the two *n-coding* CGPS groups (nCodMed and nCodHi) did not differ statistically from each other (p=0.740) although both in combination did significantly differ (p=0.003) from the conventional CGPS (nCodLo) group. Finally, all CGPS groups were significantly different from the control group (p=0.0497), thus replicating the effectiveness of Interactive Engagement (IE) methods over traditional (T) instruction[26] with novel measures.

*Does the addition of n-coding to CGPS make a significant difference in learning outcomes?*

Our results showed that under the combined effect of the multiple variables, adding *n-coding* does make a significant difference in learning outcomes. Traditionally, when significant differences between groups occur, the ANOVA difference between groups for each dependent measure is presented in order to determine what specific variables can account for the treatment differences. The ANOVA differences between groups for each variable are presented in table 4.

**Table 4: Difference Between groups under each dependent variable**

|  | nCodHi&Med vs nCodLo |
|---|---|
| *ANOVA*: FCI *g* | *p=0.115* |
| *ANOVA*: wFCI *g* | *p=0.090* |
| *ANOVA*: Exam | **p=<0.001** |
| *ANOVA*: RConf *g* | p=0.383 |
| **MANOVA** | ***p=0.003*** |

Given the small sample sizes, significant differences between *n-coding* groups and traditional CGPS were only found in final exam grades. However, contributing effects in the required direction were given by the FCI and wFCI at confidence levels close to the conventional (p<0.05) level of confidence. Note that a MANOVA could yield no significant differences under any single variable but yield significant differences when these variables are combined. Furthermore, the converse is also true: significant ANOVA differences can be found under single variables which later cancel out in the MANOVA as exemplified by differences found between high and medium *n-coding* groups in ANOVA which did not result in an overall MANOVA difference (see table 3). Regarding the difference between both *n-coding* CGPS groups and their conventional analog, these data show that the combination of statistical tendencies and significant effects on the ANOVA yield overall a highly significant multivariate difference. Therefore, the addition of *n-coding* to the CGPS approach is a valid and useful one.

Regarding the difference between the high and medium *n-coding* groups, no significant differences were found in the MANOVA (p=0.740). Although differences were detected on the underlying ANOVAs, these shall not be reported as the combined effects of the individual variables cancel out. The somewhat surprising conclusion is that no difference in learning outcomes exists between the high and medium *n-coding* groups.

*Revisiting the difference between Interactive Engagement and Traditional instruction*

Taking all interactive engagement CGPS groups (nCodHi, nCodMed & nCodLo) and comparing the outcomes with traditional instruction yields significant differences (p=0.0497) as was expected from previous findings[26]. Interestingly, the difference between Interactive engagement (IE) and Traditional (T) instruction do not only reside in FCI gains alone as previously reported[26]. The differences between IE and T groups under the four dependent measures in presented in table 5.

**Table 5: IE vs T groups under each dependent variable**

|  | nCodHi&Med&Lo vs Cont |
|---|---|
| *ANOVA*: FCI *g* | **p=0.004** |
| *ANOVA*: wFCI *g* | **p=0.001** |
| *ANOVA*: Exam | p=0.217 |
| *ANOVA*: RConf *g* | *p=0.062* |
| **MANOVA** | **p=*0.0497*** |

These data show that besides expected differences in FCI gains (p=0.004), significant differences were also found in weighted FCI scores (p=0.001) and a tendency in right answer confidence (p=0.062) is also observed. Furthermore, IE groups performed as well (or more precisely, non-significantly better) than T groups in traditional problem solving.

*Confidence data*

The last data remaining are those concerning student confidences in concepts. These data are presented in tables 6, 7 and 8.

**Table 6: Overall Average Confidence for each group**

|  | nCodHi | nCodMed | nCodLo | Control |
|---|---|---|---|---|
| Pre AvgConf | 2.8 | 2.4 | 2.4 | 2.1 |
| Post AvgConf | 3.1 | 2.9 | 2.7 | 2.3 |
| Raw Gain | 0.3 | 0.55 | 0.36 | 0.2 |
| *<g>* | 0.25 | 0.30 | 0.24 | **0.04** |
| G | **0.23** | **0.33** | **0.22** | **0.10** |

**Table 7: Average Right Answer Confidence for each group**

|  | nCodHi | nCodMed | nCodLo | Control |
|---|---|---|---|---|
| Pre RConf | 3.0 | 2.6 | 2.6 | 2.3 |
| Post RConf | 3.1 | 3.0 | 2.8 | 2.6 |
| Raw Gain | 0.17 | 0.4 | 0.2 | 0.3 |
| *<g>* | 0.21 | 0.22 | 0.09 | **0.03** |
| G | **0.16** | **0.27** | **0.16** | **0.15** |

**Table 8: Average Wrong Answer Confidence for each group**

|  | nCodHi | nCodMed | nCodLo | Control |
|---|---|---|---|---|
| Pre WConf | 2.6 | 2.2 | 2.1 | **2.1** |
| Post WConf | 2.8 | 2.7 | 2.5 | **2.2** |
| Raw Gain | 0.2 | 0.4 | 0.4 | **0.1** |
| *<g>* | 0.04 | 0.23 | 0.21 | **-0.03** |
| G | 0.14 | 0.23 | 0.20 | 0.05 |

Regardless of the section, the average overall confidence, the average right answer confidence and the average wrong answer confidence increased after instruction. This can be expected as students may feel more confident about concepts after instruction than they did before instruction. Given the small sample sizes, no significant differences were found in normalized confidence gains between any groups for these confidence data. However, although no significant differences were found in pre-test confidences between any groups (using 2-tailed paired t-tests with Bonferroni correction to adjust for the large number of tests), differences were found in post-test average confidence with the control group. Specifically, the high n-coding and medium n-coding groups differed in post-test average confidence from the control (p=0.036 and p=0.043 respectively). To find the source of this difference in average confidence, the right answer confidence and the wrong answer confidence were analyzed. No statistical differences were found between groups in average right answer confidence before or after instruction. However, although no statistical differences were found before instruction in wrong answer confidence, students in the high n-coding group were significantly more confident of their wrong answers after instruction when compared to the control group (p=0.042). Furthermore, taken together, all n-coding groups were statistically more confident of their wrong answers after instruction when compared to the control group (p=0.016).

*Discussion*

Having identified *n-coding* as a construct responsible for multiple internal representations, we set out to measure the effect of adding *n-coding* opportunities in a Cooperative Group Problem Solving (CGPS) environments. Quite consistent with the developed theoretical framework is the finding that *n-coding* CGPS is significantly different (p= 0.003) from its conventional analog. This result shows that **n-coding is an effective construct in promoting conceptual change**.

The second significant difference found was between all CGPS groups and the traditional control section (p=0.0497). These findings replicate meta-analytic differences found between traditional instruction and Interactive Engagement[26]. However, the meta-analytic finding reported primarily normalized FCI gain differences. Here, this difference was found to extend to confidence weighted FCI gains and a tendency in right answer confidence was also found. This result suggests that affective gains in confidence accompany the cognitive gains made by students in non traditional instruction. Furthermore, although more emphasis was given to concepts, students in all CGPS groups did not differ from the control group (p=0.217) in traditional problem solving. Taken together, these findings confirm that IE methods are significantly better than T methods in fostering learning, and that this enhancement goes beyond changes in Boolean conceptual states as measured by FCI scores.

Finally, no significant difference was found between the high and medium *n-coding* groups. Recall that the difference between the two *n-coding* CGPS groups (nCodMed and nCodHi) was that although both required activation of multimodal input (verbal, visual, kinesthetic, social, logical) for problem solution, only one (nCodHi) also had all these multi-modal stimuli in its presentation (see table 1). It was expected that as the *n-coding* opportunities increase, the outcome gains should also increase. The remaining question is why did the two *n-coding* groups not differ from one another?

To better understand why no difference exists, the previous question can be reformulated as: Why doesn't the introduction of multiple representations in the problem ***presentation*** contribute significantly? Constructivism may hold the answer to this question. Essentially, constructivism is the theory holding that students construct new knowledge from existent knowledge. As well put by Resnick and Hall[38] constructivism:

> *"confirms Piaget's claim that people must \*construct\* their understanding; they do not simply register what the world shows or tells them, as a camera or a tape recorder does. To "know" something, indeed even to memorize effectively, people must **build** a mental representation that imposes order and coherence on experience and information. Learning is interpretive and inferential; it involves active processes of reasoning and a kind of 'talking back' to the world - not just taking it as it comes".*

Having multimodal information necessary for task completion allows students to construct multiple representations along modalities that are not usually activated, and use these representations to "*talk back to the world*". However, giving multiple representations to students in the problem **presentation** may in fact contribute only minimally in constructing their own representation. Indeed, the effectiveness of "handing out" representations in the problem presentation runs counter to the notion that students ought to construct their own representation.

The advantage of this negative result is that the implementation of *n-coding* strategies need not be as laborious as previously thought. Implementing emulated life-like environments is not straight forward. Entire classroom settings need to be reorganized and physical props need to be placed in meaningful arrangements. Results of this study show that the processing of information (*n-coding* tasks) is more important than simulating a life-like environment. This reduces to being conscious of a variety of props and settings that make abstract problems life-like and meaningful. Many teachers do use props. This study puts an emphasis on the nature of life-like learning by specifying which additional dimensions (social, verbal, visual, kinesthetic, etc.) need to be attended to when using props in designing ill-structured instructional problems.

Concerning the confidence data it is interesting to note that the high and medium n-coding groups differed in overall confidence after instruction when compared to the control group ($p=0.036$ and $p=0.043$). It is somewhat encouraging that students following these active engagement methods seem less intimidated and display more confidence with respect to physics concepts. The most surprising finding however is the significant increase in wrong answer confidence in IE groups when compared to the control section ($p=0.016$). Why should IE students be on average more confident about misconceptions after instruction than they were before?

It may well be that students in the IE groups felt generally more confident about physics concepts and that some of this increase in confidence may have transpired to misconceptions. Furthermore, since less misconceptions were present after instruction (given that average FCI scores went up) it may well be those remaining after instruction were those most confidently held before instruction. That is, in the IE section only the most robust misconceptions persisted after a semester of active engagement instruction while weaker misconceptions vanished. Therefore, the average confidence associated to the few strong misconceptions remaining appears higher than the average confidence in all initial misconceptions. This suggests that using weighted FCI scores before instruction may be useful in identifying which misconceptions are confidently held by most students before instruction. Identifying these misconceptions at the beginning of the semester may be critical in addressing them explicitly and promoting substantial conceptual change.

*Limitations*

The results of this study and their generalizabiltiy are affected by a number of limitations. First, since the sample sizes of these groups were relatively small, replications of this study would be needed to further validate the results. Second, within the study design, 3 treatment sections and 1 control group were studied. All groups were taught by the same instructor (NL). This configuration was chosen in order to minimize "inter-instructor" differences. However, acting as one's own control can also be seen as a limitation. Indeed, much time and effort was invested in creating a range of meaningful activities. The instructor and primary investigator's enthusiasm during the treatment sections probably cannot be compared to his involvement in the "cookbook" control section. On the other hand, involving a different instructor would have been a limitation at least as severe. Variables in educational research contexts as extremely difficult to isolate and this concern is but one example of why research in education is "*not rocket science, it's much harder*"[39]. A mitigating factor however is the distance instructors have in CGPS approaches. Indeed, as a student-centered activity, it can be argued that the real involvement in the CGPS sections was that of a "*guide on the side*", leaving learners construct knowledge and negotiate meaning between themselves rather than that of a "*sage on the stage*" that directs and controls each step of the instructional process.

Third, this study takes as a starting assumption that guided inquiry is a culturally and socially acceptable instructional strategy. Where instruction is culturally expected to be formal and didactic to be effective, such as in Singapore[40], such active engagement approaches may not succeed. Cultural expectation of formal and didactic learning is also inherent in western college settings, particularly as students advance in their program. Indeed, the success of this study is not unrelated to the fact that the participating students were mostly first semester students unaware of standard college instructional formats. Had these students been finishing students who had been accustomed to formal and didactic college teaching, they may have been less likely to embrace the CGPS formats as these place different constraints on students. Since student involvement is a central part of any interactive engagement method, this issue may limit the generalizability of these findings to introductory courses.

*Conclusion*

Information presented to problem solvers in the everyday world comes in different forms. To solve an everyday problem we usually get information by looking, listening, touching interacting with people etc. Thus, it seems reasonable to believe that human problem solving can be optimized by allowing multiple modes of information to be represented and processed in parallel. The present paper has shown that adding multiple internal representations, labeled *n-coding*, is an effective strategy to enhance learning. To assess this effectiveness, new measures were developed and make use of student's confidence in concepts to better portray their knowledge states. Thus, beyond finding cognitive differences measured by FCI conceptual gains, affective differences were also found to be associated with the change of confidence students had in concepts. The implication is twofold. First, a richer non-Boolean picture of students' conceptions can be derived using confidence weighted FCI scores and other confidence data such as the change in average confidence or the change in confidence for specifically right or wrong answers. Second, using these measures, our model of internal mental representations, labeled *n-coding,* has a positive effect on learning both at the cognitive and affective levels. Thus, teachers wanting to enhance the learning of their students may opt for using ill-structured context rich problems[25] where the information in a laboratory environment can be represented in multiple ways and where the data measured from various everyday objects serve to solve the problem.